\documentclass[aps,floatfix,prc,twocolumn,showpacs,showkeys,amsmath,amssymb,superscriptaddress]{revtex4}
\usepackage{graphicx}
\usepackage{dcolumn}% Align table columns on decimal point
\usepackage{bm}% bold math
\usepackage{color}
\def\pp{$p-p$~}
\def\sr17{$\sqrt{s}$~=~17~GeV~}
\begin{document}
\title{Particle production in \pp collisions at \sr17 within the statistical model}
\author{I.~Kraus}
\affiliation{Institut f\"ur Kernphysik, Darmstadt University of Technology, D-64289 Darmstadt, Germany}
\author{J.~Cleymans}
\affiliation{Institut f\"ur Kernphysik, Darmstadt University of Technology, D-64289 Darmstadt, Germany}
\affiliation{UCT-CERN Research Centre and Department  of  Physics,\\ University of Cape Town, Rondebosch 7701, South Africa}
\author{H.~Oeschler}
\affiliation{Institut f\"ur Kernphysik, Darmstadt University of Technology, D-64289 Darmstadt, Germany}
\author{K.~Redlich}
\affiliation{Institute of Theoretical Physics, University of Wroc\l aw, Pl-45204 Wroc\l aw, Poland}
\affiliation{GSI Hemholtzzentrum f\"ur Schwerionenforschung, D-64291 Darmstadt, Germany}
\affiliation{ExtreMe Matter Institue EMMI, GSI, D-64291 Darmstadt, Germany}
\date{\today}
\begin{abstract}
A thermal-model analysis of particle production of \pp collisions at \sr17 using the latest available data is presented. The sensitivity of model parameters on data selections and model assumptions is studied. The system-size dependence of thermal parameters and recent differences in the statistical model analysis  of \pp collisions at the super proton synchrotron (SPS) are discussed. It is shown that the temperature and strangeness undersaturation factor depend strongly on kaon yields which at present are still not well known experimentally. It is conclude, that within the presently available data at the SPS it is rather unlikely that the temperature in \pp collisions exceeds significantly that expected in central collisions of heavy ions at the same energy.
\end{abstract}
\maketitle
\section{\label{secIntroduction}Introduction}
The statistical model has been used to describe particle production in high-energy collisions for more than half a century~\cite{fermiheisenberghagedorn}. In this period it has evolved into a very useful and successful model describing a large variety of data,   in particular,   hadron yields in central heavy-ion collisions~\cite{pbpb,overview} have been described in a very systematic and appealing way unmatched by any other model. It has also provided a very useful framework for the centrality~\cite{centrality} and system-size dependence~\cite{syssize,bec} of particle production. The applicability of the model in small systems like \pp~\cite{pp1} and $e^+-e^-$ annihilation~\cite{ee1} has been  the subject of several  recent publications \cite{ee2,ee3,pppred}.

The statistical-model analysis of elementary particle interactions can be summarized by the statement that the thermal parameters show almost no energy dependence in the range of $\sqrt{s}$~=14~--~900~GeV  with the temperature being about 165 MeV and the strangeness undersaturation factor $\gamma_S$  being  in the range between 0.5 and 0.7.

In the context of  the system-size dependence of particle production,  the \pp collisions at \sr17 have been  analyzed in detail recently. Based on similar data sets, the extracted parameters in different publications  deviated significantly from each other: in a previous analysis (Ref.~\cite{syssize,pppred}) we derived $T$~=~164~$\pm$~9~MeV and  $\gamma_S$~=~0.67~$\pm$~0.07, with $\chi^2/n$~=~1.7/3, while the authors in Ref.~\cite{bec} obtained $T$~=~178~$\pm$~6~MeV, $\gamma_S$~=~0.45~$\pm$~0.02 with $\chi^2/n$~=~11/7. These findings motivated different conclusions: In Ref.~\cite{syssize} no system size dependence of the thermal parameters was found, except for $\gamma_S$ which  tends to increase when more nucleons participe in the collisions but this rise is weaker than the errors on the strangeness suppression parameter. In Ref.~\cite{syssize} it was therefore concluded that the hadron gas produced in  central collisions at \sr17 reaches its limiting temperature. Based on Ref.~\cite{bec} on the other hand, it was argued in Ref.~\cite{na61} that, decreasing $\gamma_S$ and, in particular, increasing temperature towards smaller systems allow for probing QCD matter beyond the freeze-out curve established in Pb-Pb and Au-Au collisions~\cite{eovern,wheaton_phd}.

The goal of this paper is to understand the origin of these rather different thermal-model results  obtained in the analysis of \pp data. We use an up-to-date complete set of data and discuss the sensitivity of the thermal model parameters on their values. We present  systematic studies of  data used as inputs and the methods applied in their thermal model analysis.

The paper is organised as follows:
In Section~\ref{secData} we discuss the experimental data on which  different analysis are based.
In Section~\ref{secAnalysis} we summarize the main features of the statistical model and present the analysis of the SPS data obtained in \pp collisions.
In the final section we present our conclusions and summarize our results.
\section{\label{secData}Data}
\begin{figure}
\includegraphics[width=0.99\linewidth]{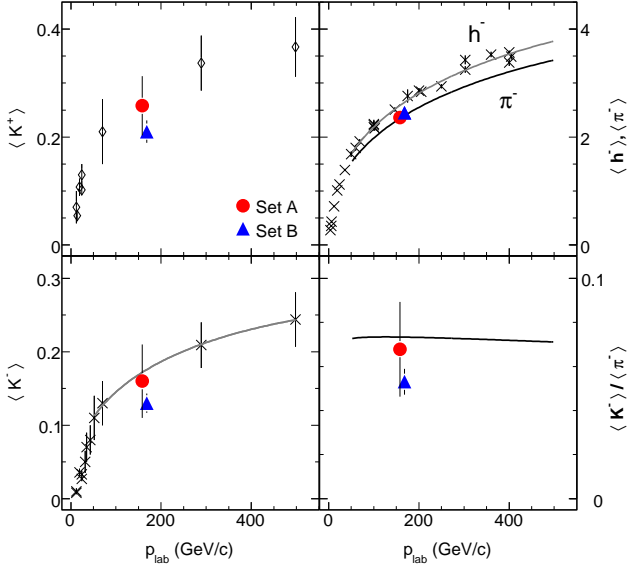}
\caption{\label{fig1} Charged kaon yields (left panels), negatively charged hadron $h^-$ and pions $\pi^-$ (upper right) and $K^-/\pi^-$
ratios (lower right panel) in \pp collisions as a function of laboratory momentum. The charged kaons, $K^+$ (diamonds) and $K^-$ (crosses) yields  are from Ref.~\cite{compK}. The lines are fits to data. The SPS yields from Ref.~\cite{dataSPSK} (circle)  and from Ref.~\cite{dataSPSK0s} (triangle) are also shown. The negatively charged hadrons are  from  Ref.~\cite{comph}.
}
\end{figure}
The data used throughout this paper for hadron yields in \pp collisions at $\sqrt s$=17.3~GeV are summarized in Table~\ref{Table_data}.
Data in column {\sl Set A} were exploit in our previous analysis~\cite{pppred} and the corresponding references are given in the table.
If the numerical values deviate in the analysis of Ref.~\cite{bec}, they are listed in column {\sl Set B}.
The relative differences between the particle yields from sets {\sl A} and {\sl B} are also indicated in Table~\ref{Table_data}.
The commonly used data in the statistical model description of particles production in \pp collisions at the SPS are displayed below the horizontal line in Table~\ref{Table_data}.
\begin{table}
% \begin{center}
\caption{\label{Table_data}
Particle yields (4$\pi$ integrated) in minimum bias \pp collisions at $\sqrt{s}$~=~17.3~GeV.
Numerical values of {\sl Set~A} are from  Ref.~\cite{pppred}. For {\sl Set~B} and common values (data  below the  horizontal line) the references are given in the last column.
}
\begin{ruledtabular}
\begin{tabular}{lccccc}
Particle & Set A & Set B & $\Delta_{yield}$ & $\Delta_{err}$ & Ref\\
\hline
$\pi^{+}$&3.02 $\pm$ 0.15 & 3.15 $\pm$ 0.16 &4.4$\%$&10.5$\%$& \cite{dataSPSpi} \\
$\pi^{-}$&2.36 $\pm$ 0.11 & 2.45 $\pm$ 0.12 &3.8$\%$&5.9$\%$& \cite{dataSPSpi} \\
$K^{+}$&0.258 $\pm$ 0.055 & &&& \cite{dataSPSK} \\
$K^{-}$&0.160 $\pm$ 0.050 & &&& \cite{dataSPSK} \\
$K^{+}$& & 0.210 $\pm$ 0.021 &19$\%$&62$\%$& \cite{dataSPSK0s} \\
$K^{-}$& & 0.130 $\pm$ 0.013 &19$\%$&74$\%$& \cite{dataSPSK0s} \\
$\Lambda$&0.116 $\pm$ 0.011 & 0.115 $\pm$ 0.012 &0.9$\%$&9.1$\%$& \cite{dataSPSlam} \\
$\bar{\Lambda}$&0.0137 $\pm$ 0.0007 & 0.0148 $\pm$ 0.0019 &8.0$\%$&171$\%$& \cite{dataSPSlam} \\
\hline
$K^{0}_{S}$& \multicolumn{2}{c}{0.18 $\pm$ 0.04} &&& \cite{dataSPSK0s} \\
$\bar{p}$& \multicolumn{2}{c}{0.0400 $\pm$ 0.0068} &&& \cite{dataSPSK0s} \\
% \hline
$\Lambda^{*}$& \multicolumn{2}{c}{0.012 $\pm$ 0.003} &&& \cite{dataSPSlamstar} \\
$\phi$& \multicolumn{2}{c}{0.0120 $\pm$ 0.0015} &&& \cite{dataSPSphi} \\
$\Xi^{-}$& \multicolumn{2}{c}{0.0031 $\pm$ 0.0003} &&& \cite{na49web} \\
$\bar{\Xi}^{+}$& \multicolumn{2}{c}{0.00092 $\pm$ 0.00009} &&& \cite{na49web} \\
$\Omega^{-}$& \multicolumn{2}{c}{0.00026 $\pm$ 0.00013} &&& \cite{na49web} \\
$\bar{\Omega}^{+}$& \multicolumn{2}{c}{0.00016 $\pm$ 0.00009} &&& \cite{na49web} \\
\end{tabular}
\end{ruledtabular}
\end{table} 

In Table~\ref{Table_set} the experimental data are grouped in sets which are used in  Section~\ref{secAnalysis} to perform the Statistical Model analysis. In the following  we motivate the particular choice of data in these sets and discuss how they can influence the model predictions on  thermal conditions in \pp collisions.
\begin{table}
% \begin{center}
\caption{\label{Table_set} Different sets of particle yields used in the thermal model fits. The type A sets contain numerical values from   the left (and common) columns of Table~\ref{Table_data}.  The type B sets contain  data from Set B  and common  columns of Table~\ref{Table_data}.}
\begin{ruledtabular}
\begin{tabular}{lll}
Set & Particles & Comment \\
\hline
A1 & $\pi^{\pm}$ $\rm K^{\pm}$ $\Lambda$ $\bar{\Lambda}$ $\rm K^0_S$ $\bar{p}$& Set of Ref.~\cite{pppred} \\
A2 & $\pi^{\pm}$ $\rm K^{\pm}$ $\Lambda$ $\bar{\Lambda}$ $\rm K^0_S$ $\bar{p}$ $\Lambda^*$ $\Xi^-$ $\bar{\Xi}^+$ $\Omega^-$ $\bar{\Omega}^+$& \\
A3 & $\pi^{\pm}$ $\rm K^{\pm}$ $\Lambda$ $\bar{\Lambda}$ $\rm K^0_S$ $\bar{p}$ $\Lambda^*$ $\phi$& \\
A4 & $\pi^{\pm}$ $\rm K^{\pm}$ $\Lambda$ $\bar{\Lambda}$ $\rm K^0_S$ $\bar{p}$& $\rm K^{\pm}$ from Ref.~\cite{dataSPSK0s} \\
\hline
B1 & $\pi^{\pm}$ $\rm K^{\pm}$ $\Lambda$ $\bar{\Lambda}$ $\rm K^0_S$ $\bar{p}$ $\Lambda^*$ &
$\Lambda^*$ contribution\\
B2 & $\pi^{\pm}$ $\rm K^{\pm}$ $\Lambda$ $\bar{\Lambda}$ $\rm K^0_S$ $\bar{p}$ $\Lambda^*$ $\Xi^-$ $\bar{\Xi}^+$ $\Omega^-$ $\bar{\Omega}^+$& Fit A of Ref.~\cite{bec} \\
B3 & $\pi^{\pm}$ $\rm K^{\pm}$ $\Lambda$ $\bar{\Lambda}$ $\rm K^0_S$ $\bar{p}$ $\Lambda^*$ $\phi$ & Fit B of Ref.~\cite{bec} \\
B4 & $\pi^{\pm}$ $\rm K^{\pm}$ $\Lambda$ $\bar{\Lambda}$ $\rm K^0_S$ $\bar{p}$ & Set B1 without $\Lambda^*$  \\

\end{tabular}
\end{ruledtabular}
\end{table}

The data set~{\sl A1} is most  restricted. Firstly, the production yields of $\Xi$ and $\Omega$ are  not included because  their numerical values are only preliminary (Ref.~\cite{na49web}). Secondly, the $\Lambda^*$ resonance is also not included  so as to restrict the analysis to  stable hadrons. Finally, the $\phi$ meson  is omitted in {\sl Set~A1} since this particle is difficult to address in the statistical model due to its hidden strangeness as discussed in  Ref.~\cite{syssize}.

The lower yields of charged kaons in {\sl Set B} of Table~\ref{Table_data} are taken from results published in conference proceedings \cite{dataSPSK0s}. Such kaon yields are in disagreement with  trends from data measured at lower and higher energies as seen  in Fig.~\ref{fig1}.

The left panels of Fig.~\ref{fig1} show the charged kaon multiplicities from \pp interactions at lower and higher beam momenta~\cite{compK} together with  data from  Table~\ref{Table_data}. The lines in this figure are simple parametrizations interpolating to  SPS energies. The $K^-$ yield from~\cite{dataSPSK} is seen to be  7\% below the expected value 
from the above parametrization, however agrees within errors. The $K^-$ abundance from~\cite{dataSPSK0s} is by 24\% lower and its error is only  10\%. As we discuss below, such a low value for the multiplicity of charged kaons influences the statistical model fit in an essential way.

The upper right panel of  Fig.~\ref{fig1}  shows the  negatively charged hadrons  from \pp interactions at several beam momenta from Ref.~\cite{comph}. As indicated in Ref.~\cite{comph}  the $K^-$, $\bar{p}$ and $\Sigma^-$ supplement the $\pi^-$ yield. In this case,  the ratio $\pi^-$/$h^-$ amounts to 91\%. Including more sources of feed-down,  the  $\pi^-$/$h^-$ ratio stays at the same level  as long as $\Lambda$ and $K^0_S$ can be separated. Consequently, to calculate the negatively charged pions from $h^-$ yields one can use the above 91\% scaling factor. Figure~\ref{fig1} (top right panel) shows the fit to $h^-$ yields as a function of  beam momenta and then by rescaling the   expected result for the $p_{\rm{lab}}$ dependence of the negatively charged pions. The yields of $\pi^-$  at SPS  from  Table~\ref{Table_data} agree quite well with that expected from an interpolation line shown in Fig.~\ref{fig1}. They are only slightly higher,  by 1\% for yields taken from \cite{dataSPSpi} and by 5\% for yields used in Ref.~\cite{bec}.

The lower right panel in Fig.~\ref{fig1} shows the $K^-$/$\pi^-$ ratio at SPS compared to the interpolated  data from other beam momenta. The mean value of the $K^-$/$\pi^-$ used in  \cite{pppred} is 8\% below the interpolated line  but  agrees within  errors, while  the corresponding value used in \cite{bec} is 28\% smaller and exhibits an error of only 11\%. Clearly, the above  differences in the $K^-$/$\pi^-$ ratios  influence the thermal model fits.

In general, a smaller  kaon yield implies a stronger suppression of the strange-particle phase space  resulting  in a smaller value for the strangeness undersaturation factor $\gamma_S$. If other strange particles are included, then  the strong suppression caused by $\gamma_S$ has to be compensated by a  higher temperature. This might be  one of the  origins for the different thermal fit parameters obtained in Refs. \cite{pppred} and \cite{bec}. In order to quantify this we have selected a data set {\sl A4} which is equivalent to the {\sl Set~A1} but with the kaon yields of Ref.~\cite{dataSPSK} being replaced by the values from Ref.~\cite{dataSPSK0s}.

The {\sl Set~B1} is (besides the $\Lambda^*$) equivalent to {\sl A1} but with numerical values for particle yields from column {\sl B} in Table~\ref{Table_data}. The {\sl Set~B4} is used to demonstrate the influence of the $\Lambda^*$ resonance on thermal fit parameters. The {\sl Sets A3}, {\sl B3} and {\sl A2}, {\sl B2} are chosen to study the influence of the $\phi$ meson and the  multistrange hyperons on thermal fit parameters.
\section{\label{secAnalysis} Statistical model analysis}
The usual form of the statistical model formulated in the grand-canonical ensemble cannot be used when either the temperature or the volume or both are small. As as a rule of thumb one needs $VT^3>1$ for a grand-canonical description to hold~\cite{hagedornred,rafeldan}. Furthermore, even if this condition is matched but the abundance of a subset of particles carrying 
a conserved charge is small, the canonical suppression still appears even though the grand-canonical description is valid for the bulk of the produced hadrons. There exists a  vast literature on the subject of canonical suppression and we refer to several articles (see e.g.~\cite{overview,polishreview,can1}).

The effect of canonical suppression in \pp collisions at ultra-relativistic energies is relevant for hadrons carrying strangeness. The larger the strangeness content of the particle, the stronger is the suppression of the hadron yield. This has been discussed in great detail in ~\cite{hamieh}.

In line with the previous statistical model studies of heavy-ion scattering at lower energies, the collisions of  small ions at SPS 
revealed~\cite{syssize} that the experimental data show stronger suppression of strange-particle yields than what  was expected in the canonical model~\cite{hotQuarks,syssize,lisbon}. Consequently, an additional suppression effect had to be included in order to quantify the observed yields. Here we introduce the off-equilibrium factor $\gamma_S\leq 1$ which reduces  densities $n_s$ of  hadrons carrying strangeness  $s$ by $n_s \rightarrow n_s \cdot \gamma_S^{|s|}$ \cite{rafeldan}.

We investigate whether or not all quantum numbers have to be conserved exactly in \pp collisions within a canonical approach by comparing data with two model settings:
\begin{enumerate}
\item [$\bullet$] {\sl Canonical (C) Model}: all conserved charges, i.e. strangeness, electric charge and baryon number are conserved
exactly within a canonical ensemble.
\item [$\bullet$] {\sl Strangeness Canonical (SC) Model}: only strangeness is conserved exactly whereas the  baryon number and electric charge are conserved on the average and their densities are controlled by the corresponding chemical potentials.
\end{enumerate}
The parameters of these models are listed in Table~\ref{Table_parameter}. In the following we compare predictions of the above statistical models with \pp data summarized in different sets discussed above.
\begin{table}
\caption{\label{Table_parameter} A list of parameters needed to quantify particle yields in the strangeness canonical (SC) and canonical (C) statistical model (see text). The symbols S, Q and B are the strangeness, electric charge and baryon number respectively with $\mu_i$ for $i=(S, Q,B)$  being chemical potentials related with conservation of these quantum numbers.
$\gamma_S$ is the  strangeness undersaturation factor, $T$ the temperature  and  $R$  the radius describing  the spherical volume of the collision zone. }
\begin{ruledtabular}
\begin{tabular}{lllll}
 & \multicolumn{2}{l}{SC model} & \multicolumn{2}{l}{C model} \\
\hline
Fit parameter		& \multicolumn{2}{l}{$\mu_B$ R} 		& \multicolumn{2}{l}{S Q R}	\\
Constrained param.	& \multicolumn{2}{l}{$\mu_Q$: B/2Q = 0.5}	& \multicolumn{2}{l}{--}	\\
Fixed param.		& \multicolumn{2}{l}{--}			& \multicolumn{2}{l}{B = 2}	\\
Fit / Scan param.	& \multicolumn{2}{l}{$T$ $\gamma_S$}		& \multicolumn{2}{l}{$T$ $\gamma_S$} \\
No. of parameter	& \multicolumn{2}{l}{5} 			& \multicolumn{2}{l}{6}	\\
\hline
			& Fit	& $\chi^2$ scan 			& Fit 	& $\chi^2$ scan	\\
No. of free param.	& 4	& 2 					& 5 	& 3	\\
No. of fixed param.	& 1 	& 3 					& 1 	& 3	\\
\end{tabular}
\end{ruledtabular}
\end{table}

\subsection{\label{secStudy} Comparative study of \pp data at SPS}
We start from the analysis of data set~{\sl A1} and modify it stepwise to find out in which way one matches the conclusion of larger  temperature in \pp than in central $A-A$ collisions at SPS as indicated in Ref.~\cite{bec}. All numerical values of model parameters are listed in Table~\ref{Table_fit}. A detailed discussion on their choice and correlations is presented in the Appendix  based on the $\chi^2/n$ systematics.
\begin{table}
\caption{\label{Table_fit} Thermal parameters extracted within the strangeness canonical (SC) and canonical (C) model (see text) from fits to 4$\pi$-integrated data in \pp collisions at $\sqrt{s}$~=~17.3~GeV.
In the SC model analysis for data set {\sl B1} a fit does not converge, the minimum of the $\chi^2$ scan is displayed.
The fits to data  sets {\sl A3} and {\sl B3} do not exhibit a $\chi^2$ minimum in the parameter range considered, thus only the tentative parameters of a possible minimum are indicated.}
\begin{ruledtabular}
\begin{tabular}{rcccccc}
\multicolumn{6}{c}{SC model results } \\
\hline
Set & $T \rm{(MeV)}$ & $\gamma_S$ & R (fm) & $\mu_B \rm{(MeV)}$ & $\chi^2/n$ \\
\hline
A1 & 163 $\pm$ 5 & 0.68 $\pm$ 0.05 & 1.50 $\pm$ 0.11 & 208 $\pm$ 14 & 1.7/4 \\
A2 & 168 $\pm$ 1 & 0.66 $\pm$ 0.02 & 1.37 $\pm$ 0.03 & 221 $\pm$ ~8 & 8.6/9 \\
A3 & $>$ 190     &  $\approx$ 0.5  & $<$ 1.1         & $>$ 250      & --    \\
A4 & 177 $\pm$ 5 & 0.59 $\pm$ 0.03 & 1.23 $\pm$ 0.10 & 233 $\pm$ 16 & 5.1/4 \vspace{1mm} \\
B1 & 176         & 0.56            & 1.24 $\pm$ 0.01 & 240 $\pm$ 12 & 7.7/7 \\
B2 & 179 $\pm$ 5 & 0.61 $\pm$ 0.02 & 1.19 $\pm$ 0.09 & 242 $\pm$ 18 &  16/9 \\
B3 &  $>$ 190    &  $\approx$ 0.5  & $<$ 1.1         & $>$ 250      & --    \\
\hline
\vspace{-2mm} \\
\multicolumn{6}{c}{C model results} \\
\hline
Set & $T \rm{(MeV)}$ & $\gamma_S$ & R (fm) & & $\chi^2/n$ \\
\hline
A1 & 175 $\pm$ 5 & 0.57 $\pm$ 0.04 & 1.33 $\pm$ 0.09 & -- & 0.5/3 \\
A2 & 174 $\pm$ 4 & 0.59 $\pm$ 0.02 & 1.34 $\pm$ 0.08 & -- & 6.6/8 \\
A3 & 189 $\pm$ 5 & 0.46 $\pm$ 0.02 & 1.12 $\pm$ 0.09 & -- &  23/5 \\
A4 & 181 $\pm$ 4 & 0.52 $\pm$ 0.03 & 1.22 $\pm$ 0.07 & -- & 3.5/3 \vspace{1mm} \\
B1 & 177 $\pm$ 5 & 0.51 $\pm$ 0.03 & 1.30 $\pm$ 0.09 & -- & 6.8/4 \\
B2 & 180 $\pm$ 4 & 0.56 $\pm$ 0.02 & 1.23 $\pm$ 0.08 & -- &  18/8 \\
B3 & 178 $\pm$ 5 & 0.45 $\pm$ 0.02 & 1.30 $\pm$ 0.10 & -- &  19/5 \\
B4 & 177 $\pm$ 5 & 0.50 $\pm$ 0.03 & 1.31 $\pm$ 0.09 & -- & 4.8/3 \\
\end{tabular}
\end{ruledtabular}
\end{table}

\begin{figure}
\includegraphics[width=0.99\linewidth]{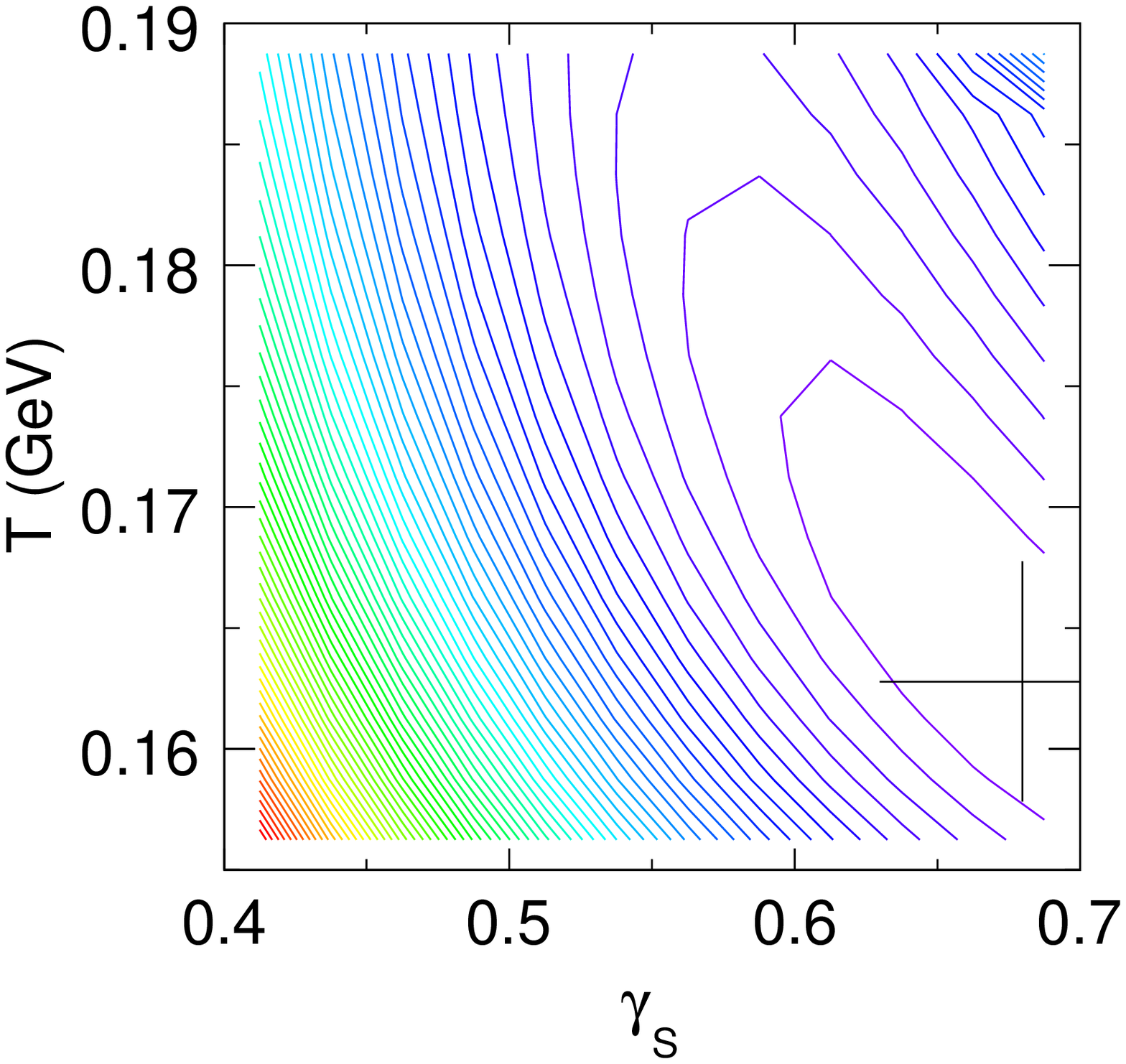}
\includegraphics[width=0.99\linewidth]{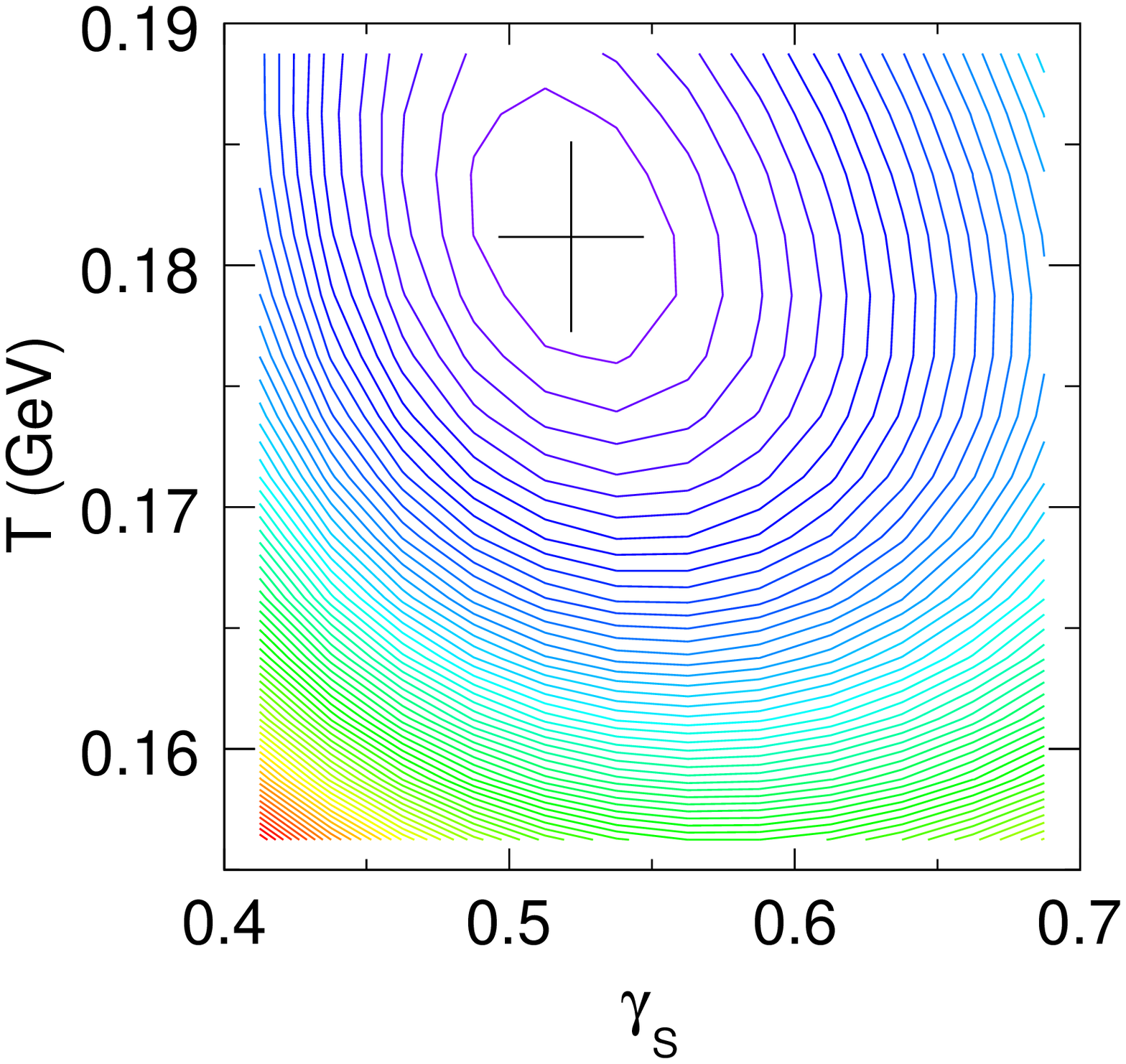}
\caption{\label{fig2}
The $\chi^2$ scan  in the (T--$\gamma_S$)-plane.
Starting from its  minimum, $\chi^2$ increases by 2 for each contour line.
Upper figure: fit to data set {\sl A1} in the model where only strangeness is conserved exactly (SC).
Lower figure: fit to data set {\sl A4} in the canonical (C) model.
The minima are indicated by the crosses.
}
\end{figure}

The fit to data  set~{\sl A1} in the SC model complies with our previous analysis from Ref.~\cite{pppred}, see also Fig.~\ref{fig2} (top). The SC model fit to these data does not  change  when including  $\Lambda^*$ hyperons resulting in the same values of thermal parameters and their errors as summarized in Table~\ref{Table_fit}.

\begin{figure}
\includegraphics[width=0.99\linewidth]{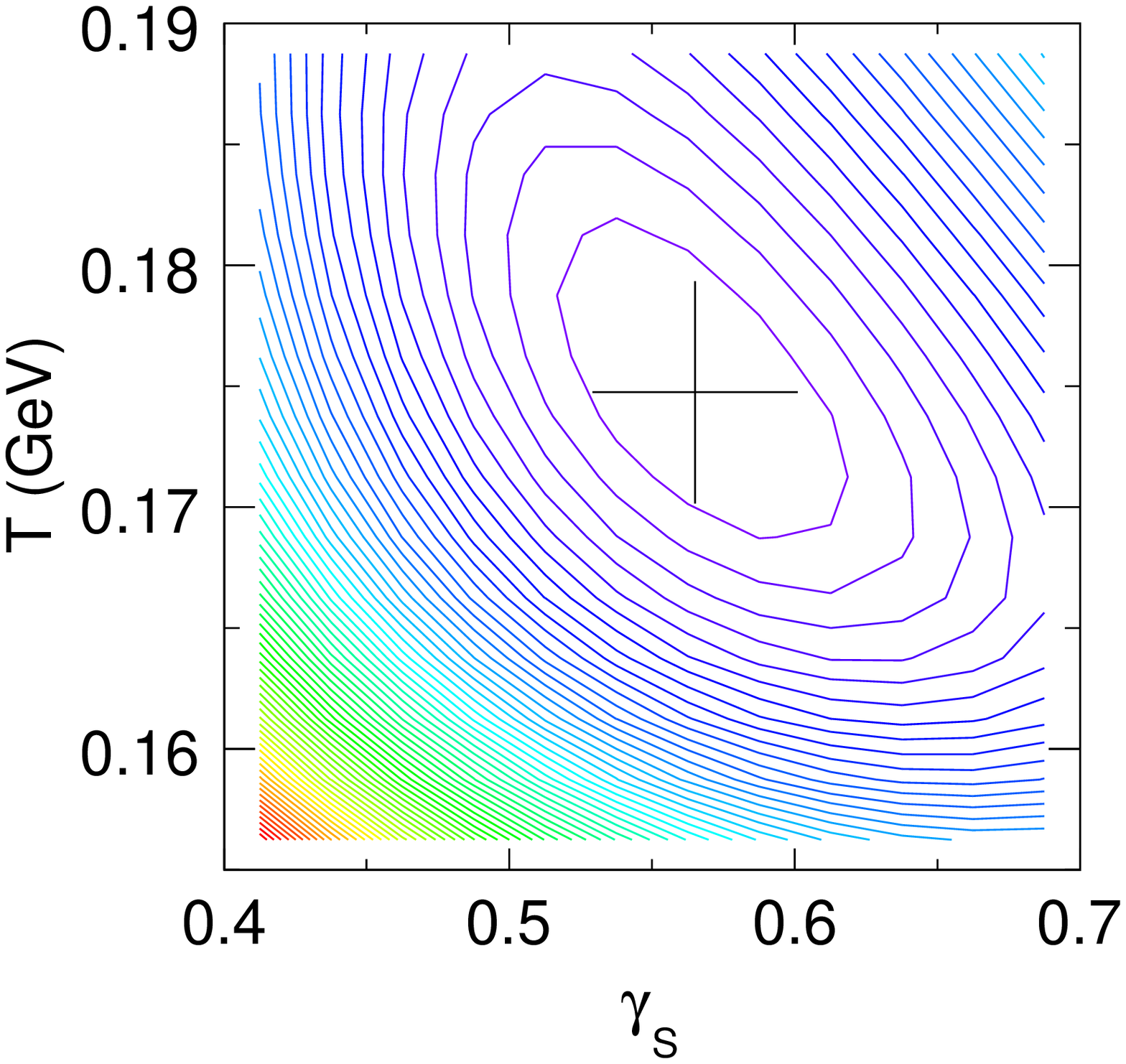}
\includegraphics[width=0.99\linewidth]{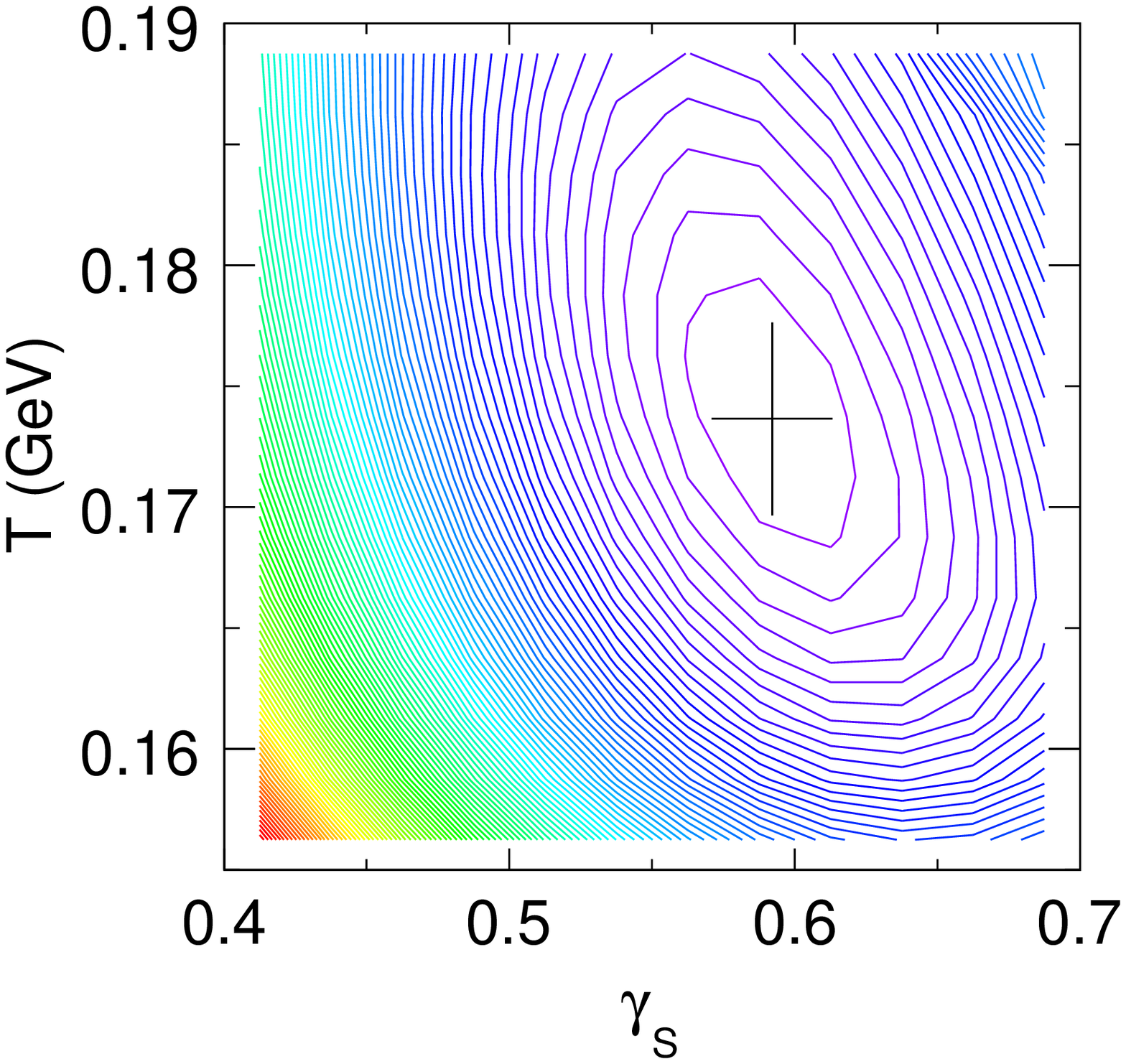}
\includegraphics[width=0.99\linewidth]{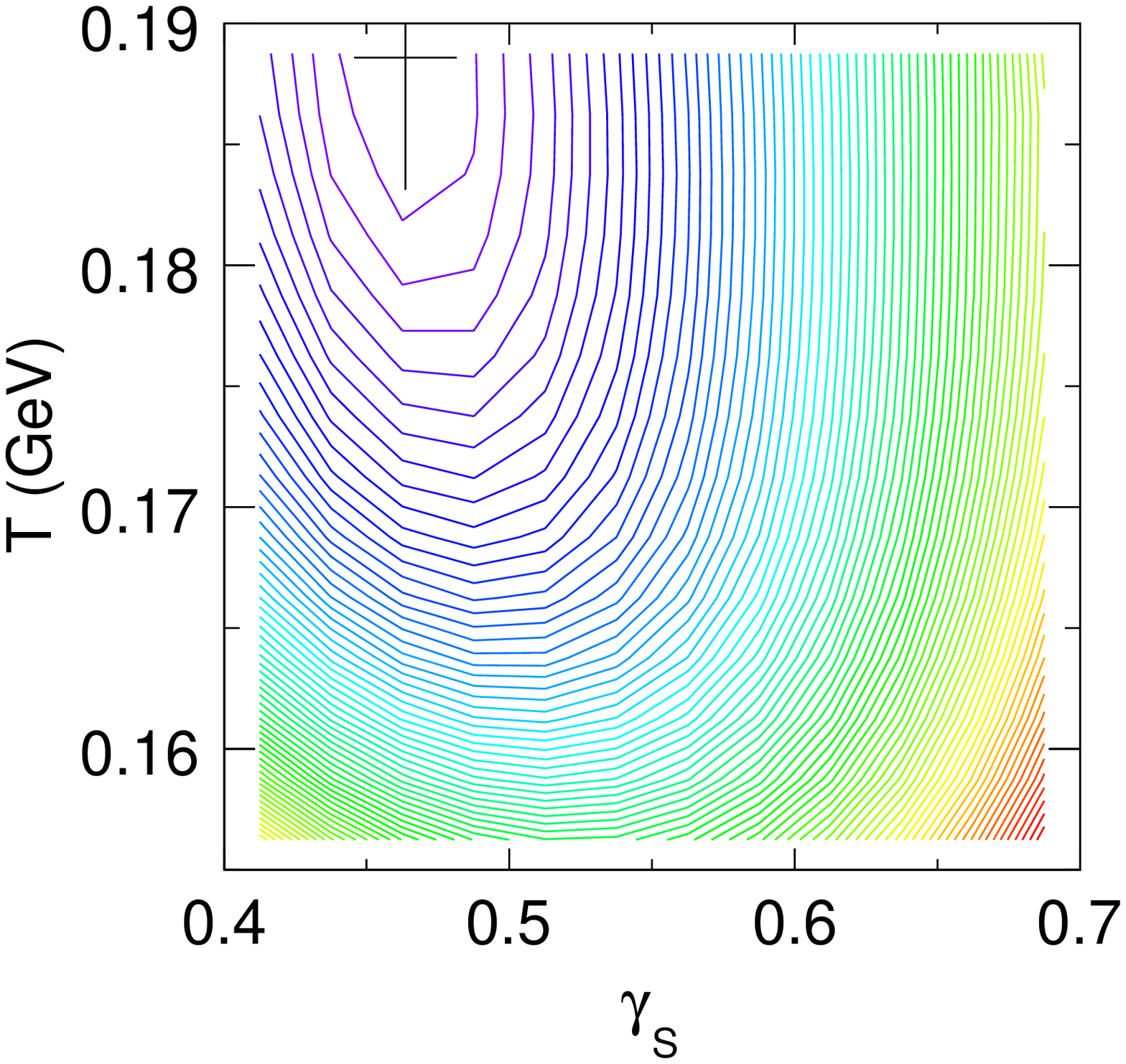}
\caption{\label{fig3} As in Fig. \ref{fig2} but for  data {\sl Set A1} (top), {\sl Set A2} (middle), {\sl Set A3} (bottom). Canonical ensemble.
}
\end{figure}

The most striking effect on thermal parameters is expected when replacing the kaon yields in {\sl Set A1} (Fig.~\ref{fig3}, top) by those from Ref.~\cite{dataSPSK0s}, {\sl Set A4}, Fig.~\ref{fig2} (bottom). Indeed, smaller kaon yields cause an increase of  temperature and a decrease of $\gamma_S$. These changes come along with a reduced volume and in case of the SC fit with increase of  the baryon chemical potential. The kaons from Ref.~\cite{dataSPSK0s} dominate the fit because their errors are 10\% while the uncertainties of the  $K^+$ and $K^-$ yields, taken from Ref.~\cite{dataSPSK}, are 21\% and 31\%, respectively. Consequently, the smaller errors dominate the statistical model fit.

In the next step we add $\Lambda^*$, $\Xi$ and $\Omega$ hyperons resulting in {\sl Set~A2}, see Fig.~\ref{fig3}, middle panel. The measured hyperon multiplicities coincide with the model results obtained before, thus within errors the statistical model parameters remain unchanged. We focus on {\sl Set~A3} and add the $\phi$ meson, Fig.~\ref{fig3} bottom. In this case  the temperature is indeed much higher. In the SC model the thermal parameters obtained from the {\sl Sets A3} and {\sl B3} appear to be meaningless and unphysical due to the $\phi$ meson contribution.

Additional to different kaon data in {\sl Sets A} and {\sl B} there are also slightly different values for pions and $\Lambda$ yields, see  Table~\ref{Table_data}. Compared to results obtained from data {\sl Sets A4}, in the fit of {\sl Set~B1} the higher pion yield reduces the strange to non-strange particle ratios resulting in slightly smaller value of $\gamma_S$. The fit obtained with the {\sl Set B4} yields similar results as that obtained from {\sl Set B1} indicating that the $\Lambda^*$ resonance does not affect the model parameters.

In the canonical (C) model analysis the {\sl Sets~A1} and {\sl A2} tend towards a slightly higher temperature and smaller $\gamma_S$ than that obtained in the SC analysis. The situation is different for {\sl Sets~A3} and {\sl B3} that include the $\phi$ meson. Here  in the SC model the temperature is very high and $\gamma_S\simeq 0.5$. In the C model the temperature decreases and $\gamma_S$ drops below 0.5. We can  conclude that in the case where the $\phi$ meson is included in the fit, one needs to apply the C analysis to get lower temperatures, however with very small values  of $\gamma_S$ and a large  $\chi^2/n$. For {\sl Set B3} the numerical results for $T$ and $\gamma_s$  summarized in Table IV coincide with that obtained in Ref~\cite{bec}, however with a larger $\chi^2/n$ \footnote{We compare our results to fit B from  Ref~\cite{bec}}.
\section{\label{secSummary} Discussion and Summary}

The statistical-model analysis of hadron yields for \pp collisions at \sr17 from Refs.~\cite{syssize} and~\cite{bec}, yield different results and lead to different conclusions on the system-size dependence of thermal parameters~\cite{syssize,na61}. In this paper we have reanalyzed the \pp data and studied the sensitivity of the thermal fit to data selection and on model assumptions. We have shown that different conclusions from Refs.~\cite{syssize} and~\cite{bec} are mostly due to differences in data selections.

Slightly different numerical values for charged pions and $\Lambda$ hyperons used in Refs.~\cite{syssize} and~\cite{bec} as well as the contribution of the $\Lambda^*$ resonance altered thermal parameters only within errors. However, the used charged kaon yields in both approaches differ substantially. We have argued that data of kaon yields in Ref.~\cite{bec} deviate from trends seen in data at different energies resulting in a higher temperature.

We have shown that higher kaon yields expected from the systematics in the energy dependence in \pp collisions are in line with data on multi-strange baryons. Unlike the hyperons, when adding the $\phi$ meson  the thermal model fit leaves  a reasonable range of parameters resulting  in a very high temperature exceeding  190~MeV and large $\chi^2/n$. We have quantified the modifications of these results when including an exact conservation of all quantum numbers in the canonical statistical model. We have shown that in the absence of $\phi$ meson the thermal fits are rather weakly influenced by canonical effects due to an exact conservation of the  baryon number and an electric charge leading in some cases to a systematic increase of the freezeout temperature. Fits including the $\phi$ meson are sensitive to an exact conservation of all quantum numbers resulting in  lower temperatures. However, the thermal model analysis of data sets with hidden strangeness has the largest  $\chi^2/n$ indicating that this particle cannot be addressed properly in this model.

From our analysis, we conclude that within the  presently available data on \pp collisions at SPS energy and  uncertainties on thermal parameters obtained from fits within the statistical model, it is rather unlikely that the temperature in \pp collisions exceeds significantly that expected in  central collisions of heavy ions at the same energy.

\begin{acknowledgments} K.R. acknowledges stimulating discussions with P. Braun-Munzinger and support of DFG, the Polish  Ministry of Science MEN, and the Alexander von Humboldt Foundation. The financial support of the BMBF, the DFG-NRF and the South Africa - Poland scientific collaborations are also gratefully acknowledged.
\end{acknowledgments}
\appendix
\section{The $\chi^2$ contours of thermal fits}
In this appendix we quantify the choice of thermal parameters within the statistical model through $\chi^2$-contours in the parameter space. Since the temperature  $T$ and $\gamma_S$ are of particular interest here the quality of the fits are shown in Figs. 2 and 3 in the $(T-\gamma_S)$-plane. In these figures for fixed $(T,\gamma_S)$-pair the remaining model parameters were fitted and the corresponding $\chi^2$ was calculated.

Figure~\ref{fig2} (top) shows the analysis of the data {\sl Set~A1} within the strangeness canonical (SC) model. The analysis in the model with canonical treatment of  all conserved charges (C) is shown in Fig.~\ref{fig3} for all data sets besides the {\sl Set A4} which is presented in Fig.~\ref{fig2} (bottom).

In the C model description of data {\sl Set~A1} there is a large region of a very low $\chi^2$ which manifests the expected anti-correlation of $T$ and $\gamma_S$. Reasonable fits are possible over a large range of parameters. For the {\sl Set~A2} the minimum is located at the same temperature and slightly higher $\gamma_S$. The contributions of $\Xi$ and $\Omega$ baryons disfavor small values of $\gamma_S$.

The $\phi$, as seen in Fig. 3, directs  fits towards  very  high temperatures and very strong strangeness suppression.  Also the pattern of ($T$--$\gamma_S$) anti-correlations shows decreasing $\chi^2$ with increasing temperature at fixed $\gamma_S$.

\end{document}